# Efficient Generation of Intense Broadband Terahertz Pulses from Quartz

Yuxuan Wei[1], Jiaming Le[1], Chuanshan Tian[1,*]

*1S Department of Physics, State Key Laboratory of Surface Physics and Key Laboratory of*

*Micro- and Nano-Photonic Structures (MOE), Fudan University; Shanghai, 200433, China.*

*Electronic mail: cstian@fudan.edu.cn

**Abstract:** The intense terahertz (THz) pulses facilitate the observation of various nonlinear optical effects and manipulation of material properties. In this work, we report a convenient approach that can produce strong broadband terahertz pulses with center frequency tunable between 2-4 THz. The coherent THz light source with pulse energy of 1.2 microjoule can be generated from a low-cost crystalline quartz pumped in the tilted wave-front scheme. Thanks to the wide transparent spectral window and high damage threshold, our theoretical analysis and experiment show that the THz conversion efficiency in quartz is comparable with that in LiNbO$_3$, and output THz spectral range is much broader. This work not only provides the light source that is urgently needed for nonlinear THz spectroscopy beyond 2 THz, but offers an alternative route in the selection of nonlinear optical crystals for optical frequency conversion.

The past decade has witnessed the unique applications of ultrafast intense terahertz (THz) radiation in probing and manipulation of material properties.[1-3] Many efforts have been made to develop an efficient, tunable, table-top terahertz emitter that produces THz pulses with a transient electric field of 1 MV/cm or larger.[4-6] Among different approaches, optical rectification (OR) of femtosecond pump pulse in $LiNbO_3$ (LN) using the tilted pulse front (TPF) scheme is currently widely adopted for the generation of intense THz source.[7,8] However, the spectral bandwidth is limited to 2 THz because of the strong absorption and dispersion in LN that greatly reduces the interaction length at higher frequencies.[9] On the other hand, there are many important elementary excitations residing beyond 2 THz, for instance, phonon modes of ferroelectric crystals[10,11] and intermolecular hydrogen bond stretching resonance in macro- or bio-molecules [12,13]. It is thus highly desired to find alternative convenient means that can extend the spectral coverage of the few-cycle THz pulses.

Various approaches had been proposed for generating high-frequency THz pulses. Optical rectification or difference frequency mixing in organic crystals, such as DAST and DSTMS, has gained increasing attention owing to their large Pockels coefficients.[14-16] This method can in principle cover the spectral range from a few THz up to the mid-infrared, except for a few narrow absorption windows[17,18]. To achieve the phase matching condition, typically, a near-IR pump light with a wavelength between 1.3 μm and 1.5 μm derived from a high-power optical parametric amplifier (OPA) is required[19]. Because the beam quality and stability of the high-power near-IR pump are relatively poor, the THz pulse generated from the organic crystals is often unstable with large pulse-to-pulse fluctuation[19]. Another concern is the quality and durability of the organic crystals, which suffer from the special millimeter-scale irregularity, deliquescence in air, and degradation caused by the pump light[20]. Alternatively, magnetic film and air-plasma pumped by a strong femtosecond pulse may also emit energetic THz pulse with broad spectral coverage[21,22]. But the pump-to-THz conversion efficiency based on the spin Hall effect or the higher-order wave mixing process is relatively low[19]. Thus, to obtain a THz pulse with energy reaching 1 μJ, a high-energy pump laser with pulse energy of >10 mJ is required[19,23], which is not readily available in ordinary laboratories.

Traditionally, a nonlinear crystal with a large $\chi^{(2)}$ is the first choice for generating intense THz radiation via OR process. Unfortunately, very few such crystals are available yet for the high THz frequencies. In this work, we propose a different approach. As will be discussed in the following, a nonlinear crystal with a wide transparent window and high damage threshold can also be used to achieve the intense THz pulse, although its $\chi^{(2)}$ is much smaller than that of LN or organic crystals. Note that the conversion efficiency $\eta_{\text{THz}}$ is proportional to $\frac{\omega_{\text{THz}}^2 d_{\text{eff}}^2 I L_{\text{eff}}^2}{n_{\text{NIR}}^2 n_{\text{THz}}} T_{\text{THz}}$ [24], where $\omega_{\text{THz}}$ is the terahertz center frequency, $d_{\text{eff}}$ is the effective second order nonlinear coefficient, $I$ is the pump intensity, $L_{\text{eff}}$ corresponds to effective interaction length[24,25], $T_{\text{THz}}$ is the terahertz transmission rate and $n_{\text{THz}}$, $n_{\text{IR}}$ are the terahertz and infrared refractive index in the nonlinear crystal. $I$ is limited by the crystal damage threshold and $L_{\text{eff}}$ is determined by the absorption of the THz and the angular group velocity dispersion of the crystal[24]. Table 1 compares the corresponding parameters between LN and crystalline α-quartz with details given in the supporting information. The multi-shot light-induced damage threshold (LIDT) of α-quartz is found to be around 0.5 J/cm$^2$ for 35fs 800 nm pulse in this experiment, which is consistent with that reported in literature[26,27], much larger than the saturation intensity of LiNbO$_3$[23]. On the other hand, $L_{eff}$ is about 3.0 mm and 0.5 mm for α-quartz and LN, respectively (see supplementary information Fig.S1(a)). The longer interaction length in quartz benefits from the negligible absorption, weak dispersion (see Fig. 1(a)), and small tilted pulse front angle. Taking the pump intensity of 0.1 J/cm$^2$ and 0.025 J/cm$^2$ for quartz and LN[28], respectively, the THz field generated is expected to be comparable in the two cases. More importantly, the THz pulse obtained from quartz can cover a much broader spectral range, i.e. 0-6THz, with lower cost.

**TABLE I**: Properties comparison between LiNbO$_3$ and Quartz. (a) the light-induced damage threshold (LIDT) of LiNbO$_3$ when pumped by sub-100 fs pulse can reach 30 mJ/cm$^2$, but it will be saturated at 10mJ/cm$^2$ because of other nonlinear phenomena such as cascading effect, and multiphoton absorption[28,29].

|  | $d_{eff}$ (pm/V) | $I$ (J/cm$^2$) | $\gamma$ (degree) | $T_{THz}$ | $L_{eff}$ (mm) | $\eta_{THz}$ (a.u.) |
|---|---|---|---|---|---|---|
| LiNbO$_3$ | 168 [ref.24] | 0.01 [ref.28(a)] | 63.2 | 0.54 | 0.5 [ref.25] | 0.39 |
| Quartz | 0.3 [ref.30] | 0.25 [ref 26] | 42.4 | 0.86 | 3 | 0.37 |

To verify the above estimation, we conducted an experimental and theoretical study on the THz generation with the tilted pulse front scheme. The generation setup is sketched in Fig. 1(b). A 36 fs, 800 nm laser with pulse energy of 2.5 mJ derived from 1 kHz Ti:sapphire amplifier was used as the pump. The intensity front of the pump pulse is titled by a gold grating with groove density of 1500 lines/mm and then imaged into quartz by two sets of cylindrical lenses in both vertical(x-axis in Fig. 1(b)) and horizontal directions(z-axis). The pump beam is normally incident on the *y*-cut plane of a right angle α-quartz prism. The generated terahertz light emits vertically from a 43° tilted plane. The trigonal axis (*z*-axis) is lying in the incident plane (see Fig. 1(b)). The polarizations of the incident pump and the output terahertz beam are along *x*-direction, which allows access of the largest nonlinear element $\chi^{(2)}_{xxx}$ of quartz.

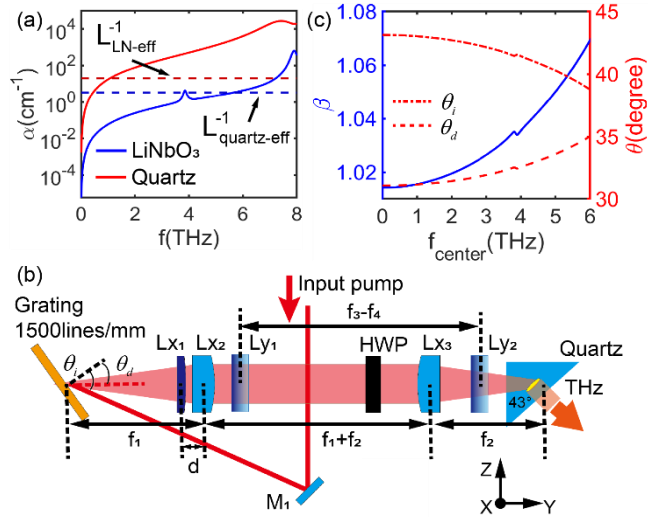

**FIG. 1.** (a) The absorption coefficient of quartz and LiNbO$_3$. (b) Sketch of the experimental setup using the titled pulse front scheme in quartz. (c) Theoretical optimized parameters versus the terahertz phase matching center frequency $f_{center}$.

$\beta = f_1 / f_2$ is the magnification factor, $\theta_i$ is the incident angle and $\theta_d$ is output angle of the grating.

The horizontal magnification factor $\beta = f_1 / f_2$ is designed to satisfied the tilted front angle that is needed for the phase-matching condition at a certain THz frequency. The focal length $f_1$ can be fine-tuned by adjusting the distance between a cylindrical lens *Lx1* with focus length of 1000 mm and an achromatic cylindrical lens *Lx2* with focus length of 250 mm. In order to minimize the walk-off effect, the pump beam is elongated along *z*-direction to 7.8 mm[24]. At the same time, to guarantee the pump intensity in the crystal, a pair of vertical cylindrical lens, *Ly1* ($f_3$ = 500 mm) and *Ly2* ($f_4$ = -50 mm), were used to focus the vertical beam diameter to 0.76 mm. It results in a pump intensity of 0.1 J/cm$^2$ for the 2.5 mJ input. The noncolinear phase-by Fourier transformation of those data in (a). (c) and (d) is |d$_{eff}$| dispersion and the calculated spectra under the three phase-matching conditions.

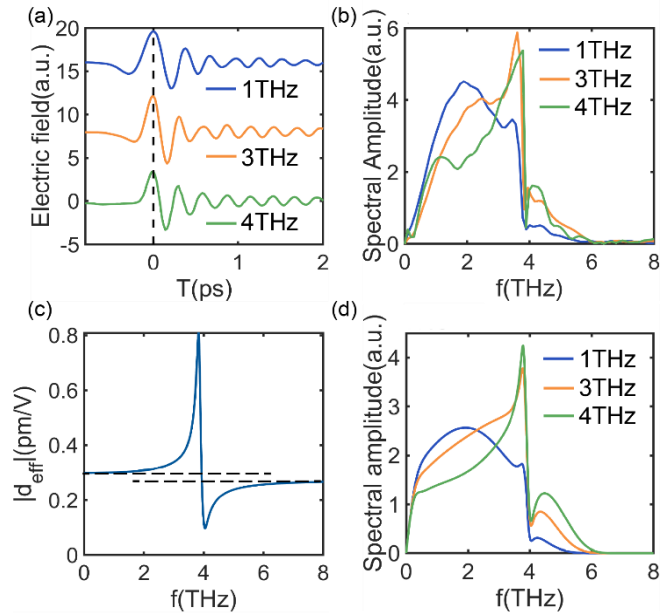

**FIG. 2.** (a) Measured temporal profile of THz pulse with phase-matching frequency optimized at 1THz, 3THz and 4THz, respectively. (b) The amplitude spectra obtained

matching condition in the tilted pulse front scheme requires $n_p^{gr}(\nu_p) = n_{THz}(\nu_{THz})\cos(\gamma)$ [25], where $n_p^{gr}$ and $n_{THz}$ are the group velocity refractive index of the pump and the phase velocity refractive index of the terahertz, respectively. The phase-matching frequency $\nu_{THz}$ can be tuned from 0 THz to 6 THz by adjusting the horizontal magnification factor $\beta$, and the incident and diffraction angles, $\theta_i$ and $\theta_d$ of grating as shown in Fig. 1(b) and (c).

Figure 2(a) presents the temporal electric field of the THz radiation generated from the quartz prism with the phase-matching center frequency $\nu_{THz}$ = 1 THz, 3 THz, 4 THz, corresponding to $\beta$ = 1.01, 1.03, 1.04, respectively. The THz fields were measured by electro-optical sampling using a 100 μm-thick GaP crystal. To avoid the absorption loss by water vapor, the THz beam path was sealed and purged with dry air. After Fourier transformation, the obtained spectra in Fig. 2(b) show clear shift of the peak position from 2THz to 4THz via tuning the phase-matching condition. The spectral coverage reaches 0-6 THz as expected, which is much wider than that generated in LN. The post-pulse oscillation with period of 0.26 ps in Fig. 2(a) and the sharp dispersive spectral feature at 3.8 THz in Fig. 2(b) originate from the lowest optical phonon mode of quartz with small absorption coefficient (see Fig. 1(a)). By considering the contribution of this resonance in $d_{eff}$ with the known resonant frequency, bandwidth, and absorption coefficient[30], the spectral features observed in the experiment can be nicely reproduced by solving the 1-D coupled-wave equations[25] as evidenced in Fig. 2(d). In the calculation, the dispersion of the nonlinear coefficient takes the form:

$$d_{eff} = d_{\infty} + \frac{(d_{\infty} - d_0)\omega_0^2}{\omega_0^2 - \Omega^2 + 2i\Omega\gamma} \quad (1)$$

where the resonant frequency $\omega_0$ =3.85 THz and the damping coefficient $\gamma$ = 0.135THz × $2\pi$[31]. The low frequency nonlinear coefficient $d_0$ is known to be 0.3pm/V from literature[30]. The high frequency nonlinear coefficient $d_{\infty}$, which is the only unknown parameter in Eq. (1), is found to be 0.27pm/V from the simulation. As shown in Fig 2(c), the severe dispersion of $|d_{eff}|$ leads to enhancement in the spectral range of 3.5-3.8 THz, but limits the spectral weight near 4 THz.

To characterize the OR efficiency in quartz, the THz energy was measured by a Far-IR power detector at various. The phase-matching condition is optimized at 3 THz. As shown in Fig. 3(a), the THz pulse energy first increases quadratically with the input pump energy as expected for a second-order optical process. Accordingly, the energy conversion efficiency scales linearly with the pump intensity. However, as the pump intensity exceeds 0.08 J/cm² (2.0 mJ/pulse), the efficiency gradually approaches saturation. The THz output can reach 1.2 µJ at the maximum pump energy of 2.5 mJ/pulse (0.1 J/cm²) available in our lab. It corresponds to energy conversion efficiency of 0.05%. Notably, the efficiency of 0.05% achieved in this experiment using a 35 fs pump pulse is comparable to LN pumped by a 100 fs pulse[7].

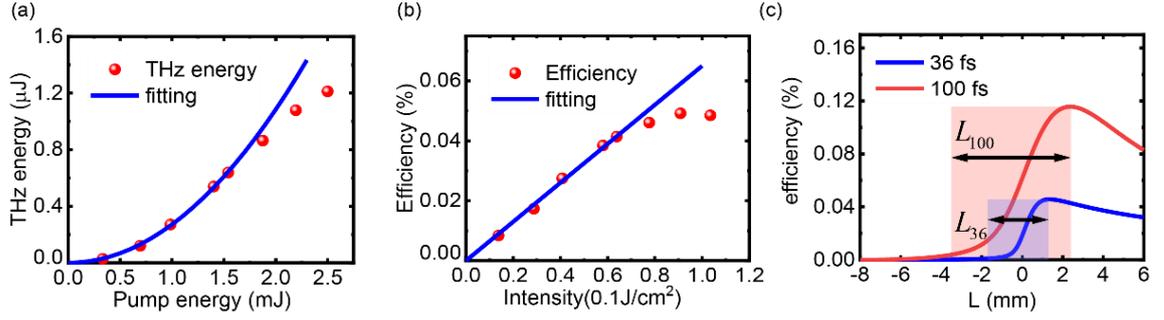

**FIG. 3.** (a) Experimental output terahertz pulse energy versus the pump energy. (b) The measured energy conversion efficiency versus the pump intensity. (c) Theoretical conversion efficiency as a function of terahertz propagation length pumped by 36 fs and 100fs 800nm pulse, respectively, with phase matching center frequency at 3THz. The shadowed area indicates the effective interaction length.

To quantitatively understand the OR process in quartz, we performed analytical calculation by taking into account the group velocity dispersion due to angular dispersion[29]. A pump intensity of 0.1 J/cm$^2$ was used for both the 36 fs pulse and the 100 fs. As shown in Fig. 3(c), the conversion efficiency at 3 THz reaches 0.045% for the 36 fs pump with effective generation length $L_{36}$ = 3 mm, but can be boosted to 0.12% if a 100 fs pulse is used. The higher efficiency in the latter case benefits from its narrower bandwidth, which results in less dispersion and longer interaction length $L_{100}$ = 6 mm. Notice that the calculated efficiency for a 36 fs pump pulse is lower than that extrapolated from the measured data in Fig. 3(b). It may be due to the value of $d_{eff}$ used in our calculation that bears large uncertain in literature [30,32] and the self-focusing effect of the pump inside quartz neglected in the calculation. It is known that the damage threshold (specified in J/cm$^2$) of the 100 fs pump is higher than the 36 fs pulse[26]. Thus, using 100 fs pump laser, one would expect to achieve THz output with pulse energy well-above 3 μJ from quartz. A transient peak THz electric field of 10 MV/cm would be realized if focused to a beam size of 0.1 mm in diameter.

In conclusion, we have reported a new method for generation of intense high-frequency terahertz pulses. Instead of choosing a nonlinear crystal with large $\chi^{(2)}$, other nonlinear crystals with high damage threshold and wide transparent window may also be efficient THz emitters. As a demonstration, we used tilted pulse front scheme to pump a wedge-shape α-quartz crystal by 2.5mJ 36fs 800nm NIR pulse with transient intensity of $0.1 J/cm^2$. The THz radiation with pulse energy up to 1.2 μJ can be obtained that covers 0-6 THz with center frequency tunable from 2-4 THz. The energy conversion efficiency up to 0.05% was obtained for the 36 fs pump, but may reach beyond 0.12% if a 100fs pump pulse is used accordingly to our calculation.

**Acknowledgement**

C.T. acknowledges the funding support from the National Natural Science Foundation of China (No. 12125403 and No. 11874123) and the Shanghai Science and Technology Committee (No. 20ZR1406000).

**Data Availability**

The data that support the findings of this study are available from the corresponding author upon reasonable request.